\documentclass{article}

     \usepackage[preprint,nonatbib]{neurips_2019}

\usepackage[utf8]{inputenc} 
\usepackage[T1]{fontenc} 
\usepackage{hyperref} 
\usepackage{url} 
\usepackage{booktabs} 
\usepackage{amsfonts} 
\usepackage{nicefrac} 
\usepackage{microtype} 
\usepackage{graphicx}
\usepackage{caption}
\usepackage{subcaption}
\usepackage{bbm}
\usepackage[ruled,vlined,noresetcount]{algorithm2e}
\SetKwComment{Comment}{$\triangleright$\ }{}

\title{Style Transfer with Time Series:\\Generating Synthetic Financial Data}

\author{%
  Brandon Da Silva\\
  OPTrust\\
  Toronto, ON M5C 3A7 \\
  \texttt{bdasilva@optrust.com} \\
   \And
  Sylvie Shang Shi\thanks{Research was done during Sylvie's co-op term at OPTrust}\\
  University of Toronto\\
  Toronto, ON M5S 2T9 \\
  \texttt{shang.shi@mail.utoronto.ca} \\
}

\begin{document}

\maketitle

\begin{abstract}
  Training deep learning models that generalize well to live deployment is a challenging problem in the financial markets. The challenge arises because of high dimensionality, limited observations, changing data distributions, and a low signal-to-noise ratio. High dimensionality can be dealt with using robust feature selection or dimensionality reduction, but limited observations often result in a model that overfits due to the large parameter space of most deep neural networks. We propose a generative model for financial time series, which allows us to train deep learning models on millions of simulated paths. We show that our generative model is able to create realistic paths that embed the underlying structure of the markets in a way stochastic processes cannot.
\end{abstract}

\section{Introduction}

In a financial markets context, "big data" often refers to the number of features one can use in their model, as opposed to the number of observations. The one exception is high frequency data, which has millions of observations per security. However, if we are designing a strategy with a longer time horizon, such as multiple days or weeks, then parameters have to be learned from a limited number of observations. As a result, many quantitative researchers reduce their feature set, at the expense of accuracy, to avoid the curse of dimensionality~\cite{verleysen2005curse}.

\begin{figure}[!htbp]
\centering
    \includegraphics[scale=0.6]{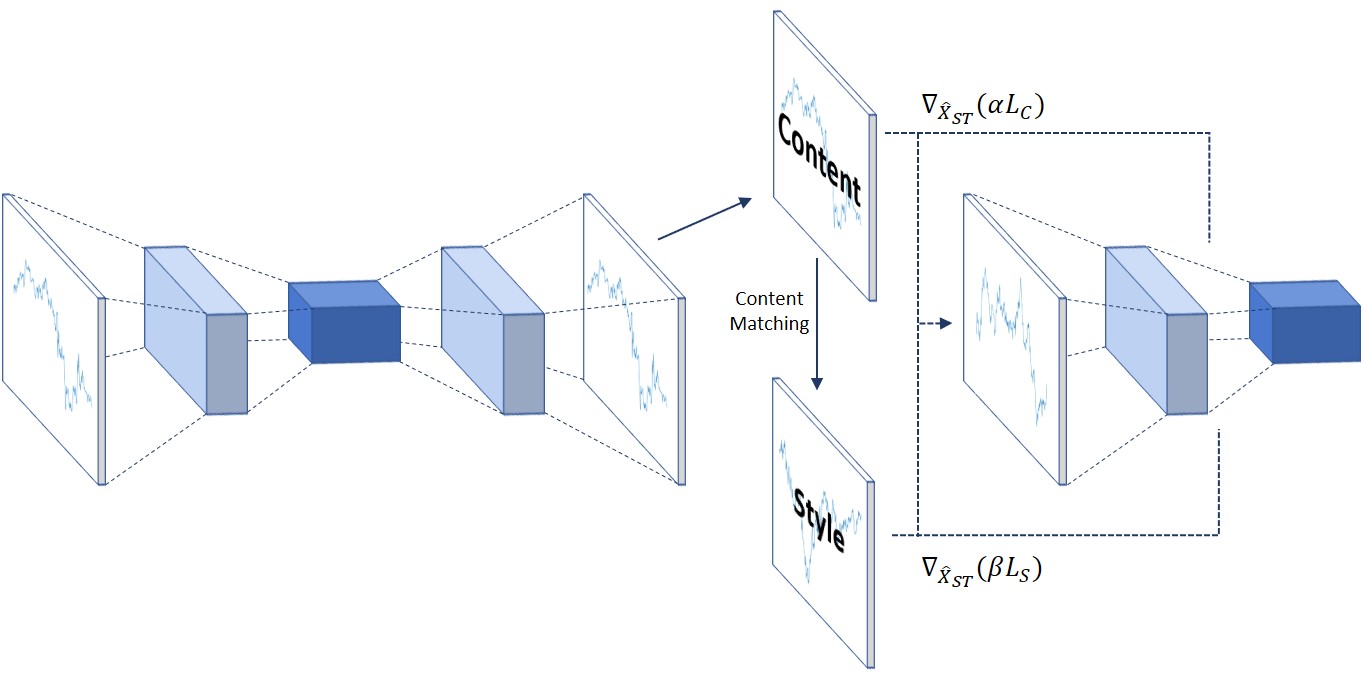}
    \caption{Although our model uses a 1D CNN, we show a 2D CNN for ease of illustration. All paths are converted to returns and normalized before going through the CNN. Corrupted returns go into the Denoising Autoencoder (DAE) to generate $\hat{X}$. The generated paths serve as the content in style transfer. Content-matching is then used to get the appropriate style paths. The gradient $\nabla_{\hat{X}_{ST}}(\alpha\mathcal{L}_C + \beta\mathcal{L}_S)$ is used to update the style transfer paths.}
\label{fig:generative-model}
\end{figure}

To increase the number of observations in a data set, other domains have used data augmentation as an effective method~\cite{perez2017effectiveness}. While popular techniques for image augmentation like cropping, flipping, and rotating are not applicable for financial time series, there are a couple papers that propose solutions for time series~\cite{le2016data, fawaz2018data}. Another way to increase the number of observations in a data set is to generate synthetic data, which is the approach taken in this paper. We use a deep generative model, which typically requires a lot of training data. As such, we take inspiration from qplum's work on recycling high frequency data~\cite{qplumHF}, and use high frequency data to train the generative model. Specifically, we use AUDUSD bid prices from May 1, 2009 - December 31, 2018, which is publicly available at \url{www.truefx.com}. Since the model generates synthetic high frequency data, we use style transfer on the generated paths, as shown in Figure~\ref{fig:generative-model}, to transfer the distributional characteristics of daily data onto our generated paths.

Historically, stochastic processes have been used as the primary method to simulate asset prices because the low signal-to-noise ratio has led many to believe that markets follow a random walk. However there are important temporal dependencies embedded in asset prices which can be observed when looking at autocorrelation plots (Figure~\ref{fig:autocorrelation-plot}). Conversely, stochastic processes have a mean autocorrelation of $\sim 0$ and a small standard deviation of $\sim 0.02$ across runs and across lags. We chose a one-dimensional convolutional neural network (CNN)~\cite{lecun1999object} as the underlying architecture for our generative model because we believe the temporal dependencies in asset prices happen locally. CNNs are great at identifying local patterns because the filters look for the same patterns everywhere in the time series, which is not the case for popular sequence modelling architectures like LSTM and GRU~\cite{hochreiter1997long, fischer2018deep, chung2014empirical}. Additionally, technical analysts believe asset prices form a hierarchy of patterns, which can be efficiently modelled with CNNs since they construct various levels of abstractions for an input. The existence of market patterns can be attributed to behavioural biases and the self-reinforcing loop that occurs because trading these patterns impacts the price of the asset being traded~\cite{murphy1999technical}.

Comparing variance estimators at different sampling frequencies was used to test the random walk hypothesis and show that stock prices do not follow a random walk~\cite{lo1988stock}. We extend this test to AUDUSD and find similar conclusions. Both the historical path and the generated paths from our model reject the null hypothesis under this specification test, while three forms of stochastic processes fail to reject the null. This indicates that our generative model captures some of the non-random local patterns that stochastic processes are not able to capture.

We believe that considering paths which have not yet happened, but retain the underlying characteristics of how asset prices move locally, will allow us to make models that generalize better. Although we are primarily concerned with the improvement it will have on deep learning, it can also be used to come up with more robust trading heuristics and better scenario-based risk modelling. The limiting factor is that inputs are restricted to price action and the features that can be derived from it, such as technical indicators.

\section{Random Walk Hypothesis}
The random walk hypothesis is a popular theory in the financial markets which states that financial price series evolve with a geometric Brownian motion (GBM)~\cite{reddy2016simulating} and are not related to historical data, which is also known as the weak form of the efficient market hypothesis. The popular stochastic pricing model relies entirely on this theory. More formally, a stochastic process $S_{t}$ is said to follow a GBM if it satisfies the following partial differential equation: 

$$dS_{t} = \mu S_{t} dt + \sigma S_{t} dW_{t}$$

where $\mu$ is the constant drift term, $\sigma$ is the volatility term and $W_{t}$ is a Wiener process or Brownian motion. By substituting $X_{t} = log(S_{t})$ and applying It\^{o}'s Lemma we can solve the above equation as:

$$X_{t} = X_{0} + (\mu - \frac{1}{2}\sigma^{2})t + \sigma \sqrt{t} \varepsilon_{t}$$

where $\varepsilon_{t}$ denotes the normal disturbance of a random walk that is identically and independently distributed. This is the strongest form of the random walk hypothesis (RW1).

In researching synthetic data generation methods, we tested the theory that time series in the financial markets do not follow random walks and thus generating prices using a stochastic model will not yield ideal results. We implemented the variance ratio test for the random walk hypothesis which debuted in Lo and MacKinlay's paper: \textit{Stock Markets Do Not Follow Random Walks: Evidence From a Simple Specification Test}~\cite{lo1988stock}. This test was made robust to the second degree of random walks, which corresponds to a GBM where the volatility of the disturbance, $\varepsilon_{t}$, is independent but not identically distributed. The heteroscedasticity includes both deterministic changes in volatility and ARCH processes in which the volatility depends on past information. This test for the random walk hypothesis is based on the fact that for two Brownian motions $B_{t}$ and $B_{s}$, the variance of the increment, $Var(B_{t} - B_{s})$, is linear in the observation interval. In other words, 

$$Var(X_{t} - X_{s}) = (t - s)Var(\varepsilon_{t}) $$

We can use this property to test for the null (RW1 and RW2) hypothesis by taking ratios of variances $\sigma_{c}^{2}$ and $\sigma_{a}^{2}$ where: 

$$\sigma_{a}^{2} = Var(X_{t} - X_{t-1}) \quad \textnormal{and} \quad 
\sigma_{c}^{2} = Var(X_{t} - X_{t - q})$$

The robustness of this test was examined by plotting the variance ratios $\sigma_{c}^{2} / \sigma_{a}^{2}$ with time series of different lengths. Variance ratios of time series generated by GBMs with changing volatility and GARCH(1, 1) both converge to unity as sample size grows indefinitely. Table~\ref{VRtest-table} shows that historical data rejects the random walk hypothesis, while classic stochastic models and volatility models fail to reject the null hypothesis. This result motivated us to continue researching for a more effective way of synthetic data generation. 

\section{Generative Model}

Although it has not been the focus for generative models, some work has been done on financial time series generation. In the related work, a heuristic-based multivariate approach is used~\cite{franco2018generating}. They focus on generating hypothetical but plausible financial time series by matching some of the observed stylized facts in the markets. While we agree with their assumption that classic models do not capture the nuances of real price action, we prefer an unsupervised approach. By using a heuristic-based approach, they are imposing a prior on the generation process, which ultimately limits the set of generated paths they can produce.

We decided to use a Denoising Autoencoder (DAE)~\cite{vincent2008extracting}, which was shown to have a probabilistic interpretation and be applied as a generative model~\cite{vincent2011connection, bengio2013generalized}, after experimenting with popular generative models such as Generative Adversarial Networks (GANs)~\cite{goodfellow2014generative} and Variational Autoencoders (VAEs)~\cite{kingma2013auto}. While the intuition behind GANs is quite elegant, they are difficult to train and often experience mode collapse (which prevent us from generalizing), vanishing gradients, and/or unstable updates~\cite{arjovsky2017towards}. Although some excellent solutions have been proposed~\cite{srivastava2017veegan,arjovsky2017wasserstein}, there is still considerable debate on which approach is best for GAN training~\cite{lucic2018gans}.

Although GANs suffer from mode collapse, their generations tend to be sharp, which contrasts the generations from VAEs, which tend to be diverse, blurry, and often unrealistic. VAEs experience these problems because the latent space from which we sample during generation is too large. Let $x$ be an input and $z$ be the transformation of $x$ onto the latent space. By extension, $q_{\phi}(z|x)$ is a probabilistic encoder parameterized by $\phi$ and $p_{\theta}(x|z)$ is a probabilistic decoder parameterized by $\theta$. For VAEs, we need to maximize the Evidence Lower Bound (ELBO):

$$\mathcal{L}(\theta,\phi,x) = \mathbbm{E}_{q_{\phi}(z|x)}[\log p_{\theta}(x|z)] - D_{KL}(q_{\phi}(z|x) \parallel p_{\theta}(z))$$

\begin{figure}[!htbp]
\centering
\begin{subfigure}[t]{0.3\linewidth}
    \centering
    \includegraphics[scale=0.6]{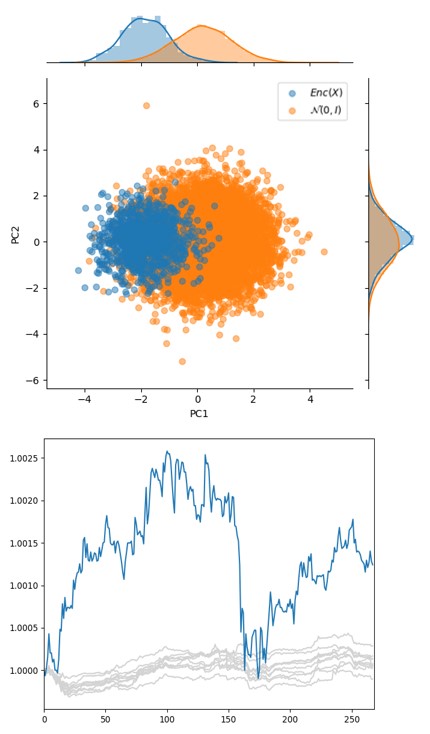}
    \caption{Regular loss (VAE)}
    \label{fig:latent-space1}
\end{subfigure}
\begin{subfigure}[t]{0.3\linewidth}
    \centering
    \includegraphics[scale=0.6]{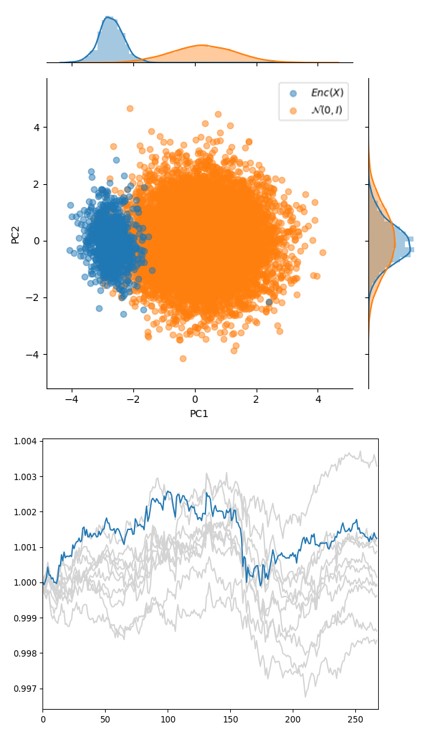}
    \caption{Overweight reconstruction loss (VAE)}
    \label{fig:latent-space2}
\end{subfigure}
\begin{subfigure}[t]{0.3\linewidth}
    \centering
    \includegraphics[scale=0.6]{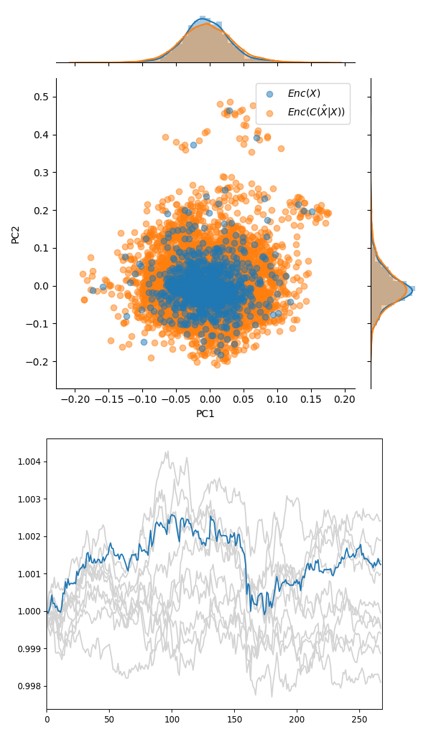}
    \caption{DAE}
    \label{fig:latent-space3}
\end{subfigure}    
\caption{(a) Using the regular loss function for VAEs, we see that reconstruction is poor, but $q_{\phi}(z|x)$ is less concentrated in $p(z)$ than (b). Placing more weight on reconstruction during training in (b), allowed for better reconstructions at the expense of sample quality during generation. (c) Using a DAE shows that the actual data distribution has high probability mass under the space that we sample from, $Enc(C(\tilde{x}|x))$, and it produces higher quality generations.}
\label{fig:latent-space}
\end{figure}

Typically, it is assumed that $p_{\theta}(z) = \mathcal{N}(0,I)$, which has empirically resulted in one of two consequences for financial time series generation. The first is that the regularization part of ELBO, $D_{KL}(q_{\phi}(z|x) \parallel p_{\theta}(z))$ is much easier to optimize than the log likelihood part, $\mathbbm{E}_{q_{\phi}(z|x)}[\log p_{\theta}(x|z)]$, resulting in poor reconstructions. An argument could be made that this is because of the low signal-to-noise ratio in financial markets. The second consequence is that observations cluster in the latent space, leaving most of the probability mass in a sub region of $\mathcal{N}(0,I)$. We show this to be true in Figure~\ref{fig:latent-space} using PCA to project the high dimensional encoding onto a 2D plane~\cite{jolliffe2011principal}. This is problematic during generation because most of our samples $z \sim p(z)$ will be non-overlapping with the actual data, $z \sim q_{\phi}(z|x)$. Some work has been done using GANs to find the part of the distribution with high probability density~\cite{makhzani2015adversarial, engel2017latent}, however we do not see the need to decouple encoding from generation. In fact, we want to use the encoder during generation to be confident that generated paths are realistic.

Let $\Omega$ define the set of paths that can be drawn from a standard normal latent space. We can also define a subset, $\mathcal{C} \subset \Omega$, where $\mathcal{C}$ represents the subset of $\mathcal{N}(0,I)$ that can generate realistic paths. Let $\mathcal{B} \subset \mathcal{C}$, in which $\mathcal{B}$ represents the subset of realistic path encodings that have been realized historically for an asset. Under the assumption that realistic paths are harder to generate than unrealistic paths, we can state that $P(\mathcal{C}) < P(\mathcal{C}^\mathsf{c})$ and $P(\mathcal{B}) \ll P(\mathcal{B}^\mathsf{c})$. 

Since we are no longer using a VAE, instead of representing our encoder as a probabilistic mapping $q_{\phi}(z|x)$, we will use a deterministic mapping $Enc(x)$. Under a DAE, we achieve variation in our model by applying a corruption process on the input $C(\tilde{x}|x) = x + \varepsilon$, where $\varepsilon \sim \mathcal{N}(0,\sigma)$. We then feed the corrupted input into the encoder to generate $z = Enc(C(\tilde{x}|x))$. Without the corruption process we can state that $X \to \mathcal{B}$, while under the corruption process $\tilde{X} \to Z$. Since the corruption process involves adding Gaussian noise to the input space, we can state that $|Z| > |\mathcal{B}|$. By keeping $\sigma$ to a small number in the corruption process, and because we assume $|\mathcal{C}| > |\mathcal{B}|$, we can be reasonably certain that $Z \subset \mathcal{C}$. Specifically, we let $\sigma$ equal to half of the realized standard deviation of the input.

It might be tempting to suggest that one can just corrupt the inputs and use that as a generative model. However, it is often the case that corrupting the inputs makes the generated paths look like stochastic process, even with a small $\sigma$, as shown in Table~\ref{VRtest-table}. On the other hand, DAEs minimize $\mathcal{L}_{DAE} = (\hat{X} - X)^2$, where $\hat{X} = Dec(Enc(C(\tilde{X}|X)))$; it learns to reconstruct the uncorrupted inputs from the corrupted inputs, which can only be the case if it learns to map $\tilde{X}$ onto the realistic portion of the latent space, $\mathcal{C}$.

Training a DAE has the added benefit of learning robust hidden representations of the underlying paths, which we use to compute content and style loss in style transfer. We use style transfer on the generated paths because our generative model is trained using high frequency data, but we want to transfer the distributional characteristics (style) of daily data onto the generated high frequency paths. Pseudo-code for the generative process is shown below: 

\vspace{2mm}

{\SetAlgoNoLine%
    \begin{algorithm}[H]
        \Indp
        \KwIn{$X$ is the set of training paths}
        \KwOut{$\hat{X}_{ST}$ is the DAE generated paths with style transfer}
        \eIf{pretained model}{
            $\theta_{DAE} \gets$ load pretrained DAE\;
        }{
            $\theta_{DAE} \gets$ initialize DAE parameters\;
            $K_{DAE} \gets$ epochs for DAE training\;
            \For{$i \gets 0$ \textbf{to} $K_{DAE}$}{
                $Z \gets Enc(C(\tilde{X}|X))$\;
                $\hat{X} \gets Dec(Z)$\;
                $\mathcal{L}_{DAE} \gets \frac{1}{n}\sum{(\hat{X} - X)^2}$\;
                $\theta_{DAE} \gets \theta_{DAE} - \eta \nabla_{\theta_{DAE}}(\mathcal{L}_{DAE})$\Comment*{Perform Adam updates for $\theta_{DAE}$}\
            }            
        }
        $\hat{X}_{ST} \gets$ initialize style transfer paths with Gaussian noise\;
        $K_{ST} \gets$ epochs for style transfer training\;
        \For{$i \gets 0$ \textbf{to} $K_{ST}$}{
            $\hat{X} \gets Dec(Enc(C(\tilde{X}|X)))$\;
            $\hat{X}_S \gets$ style paths using content-matching\;
            $F_l, \,\, P_l, \,\, S_l \gets$ feature maps in layer $l$ for $\hat{X}_{ST}$, $\hat{X}$, and $\hat{X}_S$\;
            $\mathcal{L}_C \gets \frac{1}{2}\sum{(F_l - P_l)^2}$\;
            $G_l, \,\, A_l \gets F_l^{\top}F_l, \,\, S_l^{\top}S_l$\;
            $\mathcal{L}_S \gets \sum_{l=0}^Lw_l\frac{1}{4N_l^2 M_l^2}\sum{(G_l - A_l)^2}$\;
            $\hat{X}_{ST} \gets \hat{X}_{ST} - \eta \nabla_{\hat{X}_{ST}}(\alpha \mathcal{L}_{C} + \beta \mathcal{L}_{S})$ \Comment*[R]{Perform Adam updates for $\hat{X}_{ST}$}\
        }        
        \caption{Generate financial time series}
    \end{algorithm}
}%

\section{Style Transfer}

Style transfer was originally applied to images, and allows one to transfer different textures (styles) onto an image while retaining the semantic content~\cite{gatys2016image}. We extend this framework to a 1D CNN architecture for time series. We initialize with Gaussian noise to allow the model to produce an arbitrary number of paths, as opposed to initializing with a fixed path (usually the content) that results in a deterministic mapping.

\subsection{Application for Time Series}

Style transfer has two loss functions that must simultaneously be optimized, $\mathcal{L}_C$ and $\mathcal{L}_S$. Let $F_l$, $P_l$, and $S_l$ be the feature maps for our style transfer, content, and style paths respectively. We use the last layer before the encoder as our content layer, which corresponds to the highest level of abstractions in our autoencoder. Thus by minimizing the content loss, $\mathcal{L}_C = \frac{1}{2}\sum{(F_l - P_l)^2}$, we are retaining the global paths from our DAE generations.

On the other hand, for style loss, $\mathcal{L}_S = \sum_{l=0}^Lw_l\frac{1}{4N_l^2 M_l^2}\sum{(G_l - A_l)^2}$, we use the first two layers in the autoencoder, which correspond to lower level abstractions. Let $G_l$, and $A_l$ be the Gram matrices of $F_l$ and $S_l$ respectively. The Gram matrix essentially measures the correlation between latent features, where each channel represents a feature in the hidden layer. Let us consider what the correlation between features represents in the context of a CNN. Using an RGB image as the input, we can say that when correlation between the red and blue channel is high, both colors are present, and thus a second-order feature, purple, is also present. Conversely, when correlation between the red and blue channels is 0, then purple is not present in our input image. Inputs to a CNN are essentially a combination of these second-order features since the features themselves are decompositions of the input.

Thus, by minimizing $\mathcal{L}_S$, we preserve the second-order features from our daily data. Since we use the first two layers in our DAE as style layers, we ensure that we transfer the local patterns of daily data onto our generated paths. Note that the style layers embed more granular moves in prices than the content layer because they deal with lower level abstractions. Thus, we maintain local patterns and transfer the distributional characteristics of daily data onto our high frequency generated data.

\subsection{Content-Matching}

Before applying style transfer to the generated paths, we use content-matching as a preprocessing step to select the style paths. Content-matching takes a generated path to be used as the content, and iterates through all possible style paths and selects the one with the lowest content loss. Intuitively, paths that have similar hidden representations should speed up convergence in the style transfer process because both $\mathcal{L}_C$ and $\mathcal{L}_S$ are functions of the hidden representations. Empirically, we observed this to be true. 

Dynamic Time Warping (DTW)~\cite{berndt1994using} and FastDTW~\cite{salvador2007toward} were initially explored as alternatives to content-matching. The paths that these algorithms matched were more accurate, but they do not scale well to thousands of paths. Let $n$ be the sequence length of a path, $M$ be the number of content paths, and $J$ be the number of possible style paths. Comparing one time series against another has a time complexity of $\mathcal{O}(n^2)$ and $\mathcal{O}(n (8r +14))$ for DTW and FastDTW respectively. Under the full path-matching process, in which we iterate through each content path and each style path, the time complexities go to $\mathcal{O}(JMn^2)$ and $\mathcal{O}(JMn(8r +14))$ for DTW and FastDTW respectively.

For a 1D CNN, let $K$ be the kernel size, and $S$ be the stride size. The time complexity for a given layer is $\mathcal{O}(K((n_l-K)/(S)+1))$, where $n_l$ is the sequence length in each layer. For our experiments we set $K=S$, which reduces the complexity to $\mathcal{O}(n_l)$ without a constant. If we take into account the number of channels we output in each layer $c_l$, and the number of layers $L$, the time complexity becomes $\mathcal{O}(\sum_{l=0}^{L}n_lc_l)$. For a large $n_l$ and a small $K$, $n_l \approx n_{l-1}/S$. Given we use the first two layers for content-matching, the time complexity becomes $\mathcal{O}(n(c_1 + c_2/S))$. Under the full path-matching process, the time complexity goes to $\mathcal{O}(JMn(c_1 + c_2/S))$.

Considering $n$ is usually in the hundreds or thousands, DTW is eliminated as an option due to poor scalability. Next we compare the constants for FastDTW $(8r +14)$ and content-matching, $(c_1 + c_2/S)$. Given our hyperparameters $c_1=8, c_2=16, S=3$, content-matching has a lower constant than FastDTW even with $r=1$. However, if we consider the fact that a higher $r$ is required to approach DTW, content-matching is much faster. Ultimately we chose content-matching because it achieved a similar speedup in style transfer convergence as DTW and FastDTW, but with a fraction of the computational burden.

\section{Experimental Results}

For our experiments, we first wanted to see if our model is able to generate paths that are sufficiently different from a random walk. To demonstrate this, we use the variance ratio test on historical paths, corrupted historical paths, stochastic processes, and paths generated from our model. To set up the test we show that historical paths, both high frequency and daily, reject the null hypothesis, indicating that they do not follow a random walk. To compare, 10,000 paths were constructed with a sequence length equal to that of daily data for each generative model. Table~\ref{VRtest-table} shows that the three types of stochastic processes fail the specification test, indeed confirming that this test is able to identify random walks. Interestingly, when the inputs are corrupted with Gaussian noise, even with small $\sigma$, they fail to reject the null. On the other hand, the paths that were generated with our model reject the null, indicating that they do not follow a random walk.

\begin{table}
  \caption{Variance ratio test (q=2, 95\% confidence)}
  \label{VRtest-table}
  \centering
    \begin{tabular}{llll}
        \addlinespace
        \hline \\[-1.5ex]
         Path Type &  p-Value & Reject Null \\[0.5ex] \hline \\[-1.5ex]
         Historical High Frequency & 0.000 & Reject \\    
         Historical Daily & 0.006 & Reject \\
         GBM (Constant $\sigma$) & 0.240 \,\, $\pm \, 0.140$ \,\, & Fail to Reject \\
         GBM (Stochastic $\sigma$) & 0.234 \,\, $\pm \, 0.144$ \,\, & Fail to Reject \\
         GARCH(1,1) & 0.232 \,\, $\pm \, 0.130$ \,\, & Fail to Reject \\
         Corrupted Historical Daily & 0.082 \,\, $\pm \, 0.101$ \,\, & Fail to Reject \\
         DAE with Style Transfer & 0.011 \,\, $\pm \, 0.029 $ \,\, & Reject \\
    \end{tabular}
\end{table}

\begin{figure}[!htbp]
\centering
\begin{subfigure}[t]{0.48\linewidth}
    \centering
    \includegraphics[scale=0.4]{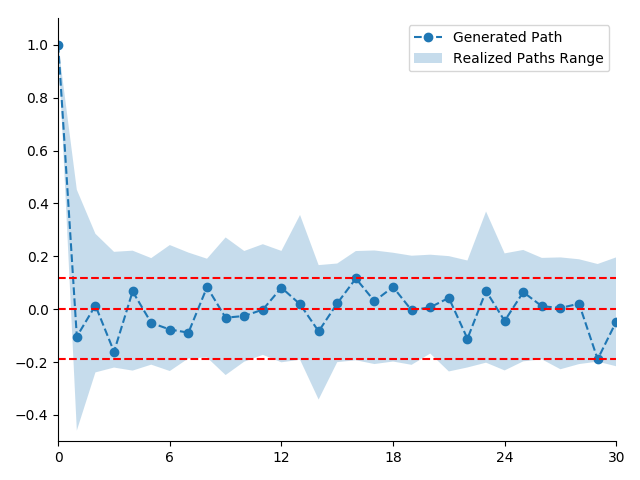}
    \caption{Autocorrelation range for high frequency data and autocorrelation for DAE generated path}
\end{subfigure}
\begin{subfigure}[t]{0.48\linewidth}
    \centering
    \includegraphics[scale=0.4]{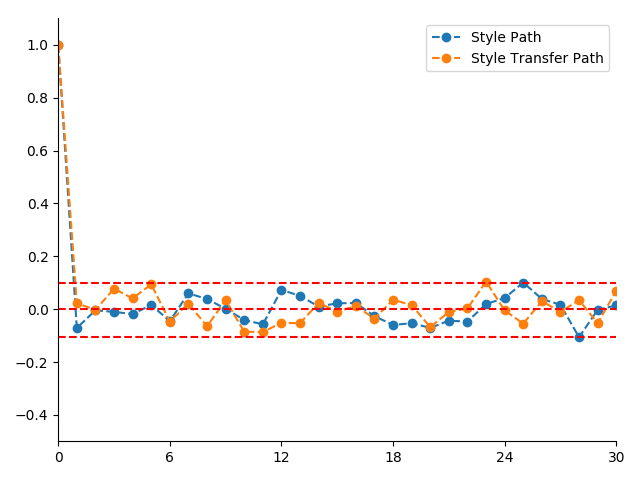}
    \caption{Autocorrelation for style transfer path}
\end{subfigure}
\caption{Autocorrelation confirms generated paths are realistic}
\label{fig:autocorrelation-plot}
\end{figure}

The variance ratio test is robust to heteroscedastic increments observed in financial time series, which means that by taking ratios of variances we are really testing the statistical significance of autocorrelations in the data~\cite{lo1988stock}. Thus, it should follow that autocorrelation is present in our model's generated paths and the realized paths, but not in stochastic processes. Across all generated paths from stochastic processes, and across 30 lags, the autocorrelation was shown to be $0 \,\, \pm 0.02$. This contrasts the autocorrelation observed in historical data (Figure~\ref{fig:autocorrelation-plot}). We show that the high frequency generated paths from our DAE lies nicely within the autocorrelation range observed across thousands of trading days. Similarly, after applying style transfer, our synthetic daily path lies within the same range as our style path (content-matched realized daily path).

Initially we thought a simple volatility scaling of returns could be appropriate to transform the high frequency paths to daily paths. But this would leave autocorrelation and higher order moments unchanged. By applying style transfer to the DAE generated paths, we are able to generate paths with similar autocorrelation, skew, and kurtosis of daily data, in addition to the mean and standard deviation (Table~\ref{higher-moment-table}).

\begin{table}
  \caption{Matching higher-order moments}
  \label{higher-moment-table}
  \centering
    \begin{tabular}{c|ccc|ccc}
          \addlinespace
         & \multicolumn{3}{c|}{One Example Path} & \multicolumn{3}{c}{All Generated Paths} \\ \hline \\[-1.5ex]
         Statistic &  Content & Style & Style Transfer & Content & Style & Style Transfer \\[0.5ex] \hline \\[-1.5ex]
         Mean & -0.00001 & -0.00010 & -0.00009 & 0.00000 & -0.00007 & -0.00009 \\    
         Standard Deviation & 0.00030 & 0.00511 & 0.00570 & 0.00041 & 0.00514 & 0.00550 \\
         Skew & -1.36 & 0.12 & 0.20 & 0.68 & 0.01 & 0.05 \\
         Kurtosis & 13.71 & 0.59 & 1.21 & 38.44 & 1.67 & 0.53 \\
    \end{tabular}
\end{table}

Furthermore, we use the Kolmogorov-Smirnov test (KS test) to compare the outputs of different generative approaches. In this case, we are testing the null hypothesis that each generated path is drawn from the same distribution as the historical daily returns. The test was evaluated on 10,000 generated paths for each approach, and the results are shown in Table~\ref{KStest-table}. The only approach that fails to reject the null is our generative approach, indicating that it is able to capture the distributional characteristics of daily data better than stochastic processes. 

\begin{table}[!htbp]
  \caption{Kolmogorov–Smirnov test (95\% confidence)}
  \label{KStest-table}
  \centering
    \begin{tabular}{llll}
        \addlinespace
        \hline \\[-1.5ex]
         Generative Approach &  \multicolumn{2}{l}{p-Value} & Reject Null \\[0.5ex] \hline \\[-1.5ex]
         GBM (Constant $\sigma$) & 0.023 & $\pm \,\, 0.046$ \,\, & Reject \\
         GBM (Stochastic $\sigma$) & 3.4e-09 & $\pm$ 1.4e-07 \,\, & Reject \\
         GARCH(1,1) & 4.4e-14 & $\pm$ 1.6e-13 \,\, & Reject \\
         DAE (High Frequency) & 8.4e-15 & $\pm$ 2.9e-13 \,\, & Reject \\
         DAE with Style Transfer & 0.571 & $\pm \,\, 0.262$ \,\, & Fail to Reject \\
    \end{tabular}
\end{table}

\begin{figure}[!htbp]
\centering
\begin{subfigure}[t]{0.3\linewidth}
    \centering
    \includegraphics[scale=0.27]{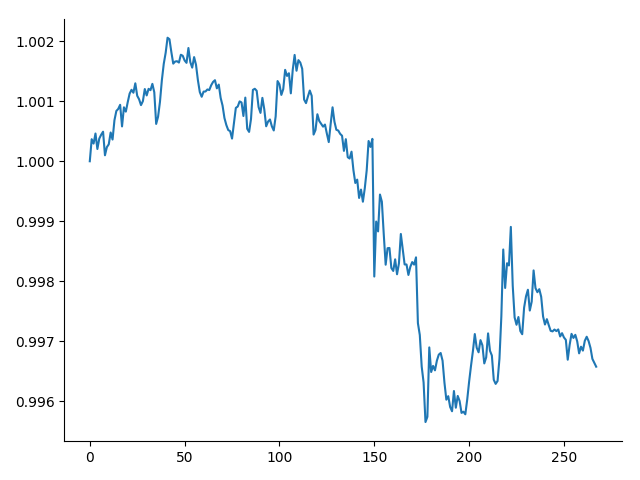}
    \caption{Content}
\end{subfigure}
\begin{subfigure}[t]{0.3\linewidth}
    \centering
    \includegraphics[scale=0.27]{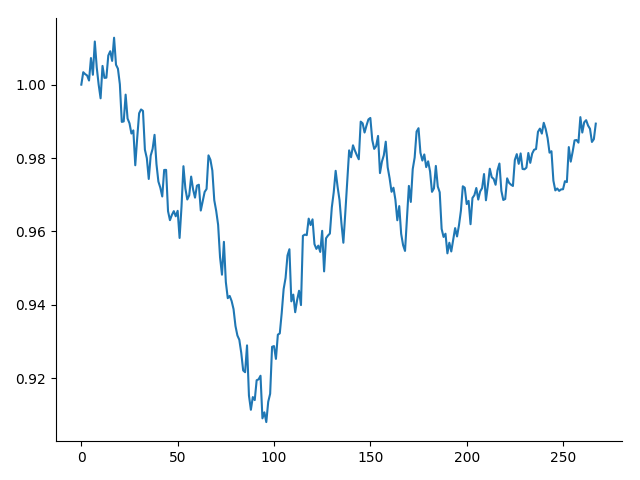}
    \caption{Style}
\end{subfigure}
\begin{subfigure}[t]{0.3\linewidth}
    \centering
    \includegraphics[scale=0.27]{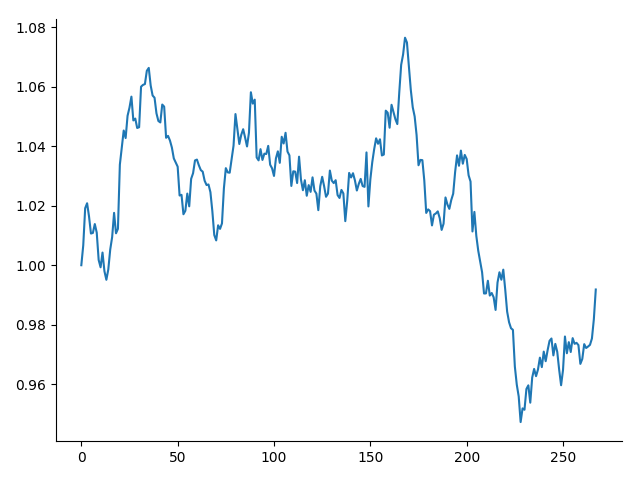}
    \caption{Style Transfer}
\end{subfigure}
\caption{We first generate a high frequency path using our DAE, which is used as the content. Then we use content-matching to find the best style path, and ultimately use both to generate (c)}
\label{fig:style-transfer}
\end{figure}

We include the high frequency generated paths in the comparison to demonstrate why we need to use style transfer. In Figure~\ref{fig:style-transfer}, we show an example of the generation process. 

\begin{figure}[!htbp]
\centering
\begin{subfigure}[t]{0.3\linewidth}
    \centering
    \includegraphics[scale=0.33]{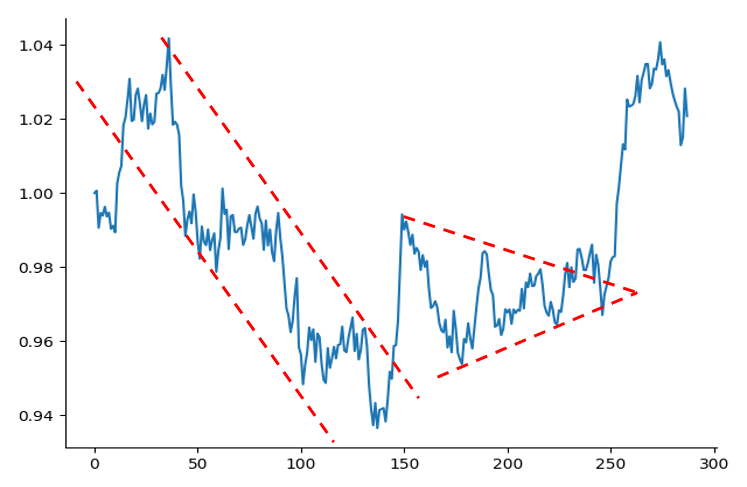}
    \caption{Generated Path}
\end{subfigure}
\begin{subfigure}[t]{0.3\linewidth}
    \centering
    \includegraphics[scale=0.33]{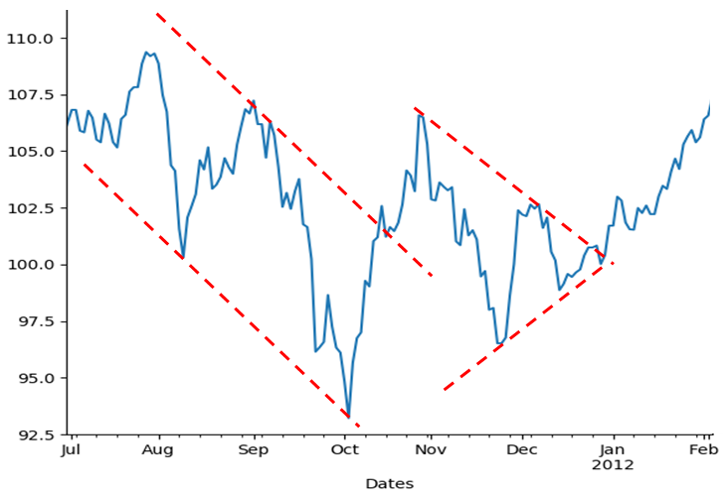}
    \caption{In-Sample Historical Path}
\end{subfigure}
\begin{subfigure}[t]{0.3\linewidth}
    \centering
    \includegraphics[scale=0.33]{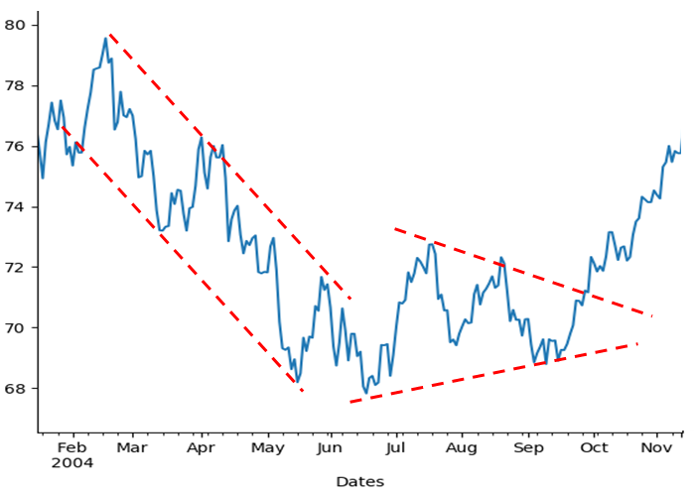}
    \caption{Out-of-Sample Historical Path}
\end{subfigure}
\caption{AUDUSD experienced similar sequences of patterns compared to the generated path}
\label{fig:technical-chart}
\end{figure}

Lastly, we conducted a visual inspection of our generated paths and were impressed to see that they embed popular technical patterns observed in asset price movements; our process generates realistic paths that embed the structure of the markets. Technical analysts will be familiar with two very common patterns shown in Figure~\ref{fig:technical-chart}: channels (not to be confused with CNN channels) and triangles. We found similar patterns in hundreds of our generated paths, but struggled to find similar patterns in paths generated with stochastic processes. Occasionally we found channels in GBM (constant $\sigma$) generated paths, but nothing as clear as our approach. Both GBM (stochastic $\sigma$) and GARCH(1,1) were unable to produce realistic technical patterns.

\section{Conclusion}

In this paper we introduced a generative model which produces realistic financial time series. The simulated paths can be used to train deep learning models, construct robust trading heuristics, and conduct more realistic scenario-based risk modelling. We demonstrated that our generated paths do not follow a random walk, and display some of the popular patterns observed in the markets. Future work will be focused on extending this framework to allow for multivariate path generation, in which the complex relationships between time series are encapsulated in the generation process.

\subsubsection*{Acknowledgments}

We thank Alex Yau and Tazeen Ajmeri for their assistance with VAE experimentation during their time at OPTrust.


\end{document}